\def\be{\begin{equation}}
\def\ee{\end{equation}}
\def\ba{\begin{array}}
\def\ea{\end{array}}

\documentclass[preprint,showpacs,amsmath]{revtex4}
\usepackage{pifont}
\usepackage{amssymb}
\def\qed{\leavevmode\unskip\penalty9999 \hbox{}\nobreak\hfill
     \quad\hbox{\leavevmode  \hbox to.77778em{%
               \hfil\vrule   \vbox to.675em%
               {\hrule width.6em\vfil\hrule}\vrule\hfil}}
     \par\vskip3pt}

\newtheorem{theorem}{Theorem}

\newtheorem{lemma}{Lemma}
\input amssym.def
\begin{document}
\title{\large\bf Local Unitary Invariants of Quantum States}
\author{ Meiyu Cui$^1$, Jingmei Chang$^1$, Ming-Jing Zhao$^{2}$, Xiaofen Huang$^{13}$,  Tinggui Zhang$^{13\dag}$\\[10pt]
\footnotesize
\small$1$ School of Mathematics and Statistics, Hainan Normal University,\\
\small Haikou 571158, P. R. China\\
\small$2$ School of Science,
Beijing Information Science and Technology University, 100192, Beijing, China.\\
 \small$3$ Hainan Center for Mathematical Research, Hainan Normal
University,\\
\small Haikou 571158, P. R. China\\
\small $^\dag$ Correspondence to
tinggui333@163.com}
\date{}

\bigskip

\begin{abstract}

We study the equivalence of mixed states under local unitary transformations in bipartite system and three partite system. First we express quantum states in Bloch representation.
Then based on the coefficient matrices, some invariants are constructed in terms of the products, trace and determinant of matrices. This method and results can be extended to multipartite high dimensional system.
\end{abstract}

 \maketitle

\bigskip

\section{Introduction }

Quantum states of a composite system can be divided into entangled
and separable. Entangled states are considered a necessary resource
for quantum communication and quantum computation \cite{nie}. Two
entangled states are said to be equivalent in implementing the same
quantum information task if they can be obtained with certainty from
each other via local operation and classical communication (LOCC).
Mathematically, this LOCC equivalent class is defined such that
within this class any two quantum states are interconvertible by
local unitary (LU) operators \cite{wgj}.

In recent years, there have been many results on the classification
of quantum states under LU. For pure bipartite states, the equivalence under LU can be done by Schmidt
decomposition \cite{nie}. For n-qubit pure
states, a
general way to determine the LU equivalence is proposed by Kraus \cite{mqubit}. Later, this method is reformed by Liu to
solve the LU classification of high dimensional multipartite
pure states \cite{bjxc}. For pure multipartite states, the classification under LU is studied by \cite{haa,fjb}.
For mixed states, a necessary
and sufficient criterion of the LU equivalence for
general multipartite states based on matrix realignment is presented in Ref. \cite{zzlf}.
In
Ref. \cite{lmzt}, a computable criterion based on Bloch
representation is presented for n-qubit mixed states. By this
method, we also considered LU equivalence of isotropic-like
states \cite{zhlz}.
Ref. \cite{ggbk} derives some necessary and sufficient conditions
for arbitrary multi-mode (pure or mixed) Gaussian states to be
equivalent under Gaussian local unitary operations. Ref. \cite{ntog} discusses the LU problem of graph
states and hypergraph states.

Another method to study the LU equivalence of quantum states is
invariant. There have many results on calculation of invariants
related to the equivalence of quantum states under LU
transformations.
In Ref. \cite{ym}, a
complete set of 18 polynomial invariants is presented for the
local unitary equivalence of two-qubit mixed states. Partial
results have been obtained for three-qubit states \cite{nsa} and some generic mixed states \cite{bs,ssx,Albeverio-Fei,Albeverio-Fei-D}.
For mixed states, Ref. \cite{ctsnx} solves the local unitary equivalence
problem of arbitrary dimensional bipartite nondegenerated
quantum systems by presenting a complete set of invariants.
Recently,
Ref. \cite{amma} derives necessary and sufficient conditions for the
LU equivalence of two  general n-qubit states using
the 1-qubit reduced states of the given multiqubit states. Ref. \cite{jnlm}
presents a complete set of local unitary invariants for generic
multi-qubit systems which gives necessary and sufficient conditions
for two states being local unitary equivalent.
These invariants are
canonical polynomial functions in terms of the generalized Bloch
representation of quantum states.

In this article we consider the LU problem for quantum
states in terms of Bloch representation in bipartite system and three partite system. We give a series of
invariants by the coefficient matrices under Bloch representation, which are necessary conditions for the LU equivalence. This method and result can be extended to multipartite high dimensional system.

\section{LU Invariants of Bipartite Quantum States }

Now, we consider bipartite systems in a $d_1 \times d_2$ dimensional
Hilbert space $H_A^{d_1} \otimes H_B^{d_2}$. Two bipartite states
$\rho$ and $\rho'$ are said to be local unitary equivalent if there
exist unitary operators $U_1\in{SU(d_1)}$, $U_2\in{SU(d_2)}$ such
that
\begin{equation}\label{ass}
\rho^\prime=(U_1\otimes{U_2})\rho(U_1\otimes{U_2})^\dag.
\end{equation}

For simplicity, we use $\lambda^1_i$, $\lambda^2_j$,
$i=1,2,\cdots,d_1^2-1$, $j=1,2,\cdots,d_2^2-1$ to denote the
generators of $SU(d_1)$ and $SU(d_2)$. In general, bipartite mixed
states $\rho$ and $\rho^\prime$ acting on $H_A^{d_1} \otimes H_B^{d_2}$ can be decomposed in
the following way:
\begin{equation}
\rho=\frac{1}{d_1d_2}I \otimes I + \sum_i S_i \lambda^1_i\otimes I +
\sum_j T_{j} I \otimes \lambda^2_j + \sum_{ij}R_{ij}\lambda^1_i
\otimes \lambda^2_j,
\end{equation}
\begin{equation}\label{rhoprime}
\rho^\prime=\frac{1}{d_1d_2}I \otimes I + \sum_i S^\prime_i \lambda^1_i\otimes I + \sum_j T^\prime_{j} I \otimes \lambda^2_j +
 \sum_{ij} R^\prime_{ij}\lambda^1_i \otimes \lambda^2_j,
\end{equation}
where $S^{(\prime)}_i=Tr(\rho^{(\prime)}\lambda^1_i\otimes I)$, $T^{(\prime)}_j=Tr(\rho^{(\prime)} I \otimes
\lambda^2_j)$, $R^{(\prime)}_{ij}=Tr(\rho^{(\prime)}\lambda^1_i \otimes \lambda^2_j)$, $S^{(\prime)}=(S^{(\prime)}_i)$ is a $d_1^2-1$ dimensional vector, $T^{(\prime)}=(T^{(\prime)}_i)$ is a $d_2^2-1$ dimensional vector, $R^{(\prime)}=(R^{(\prime)}_{ij})$ is a $(d_1^2-1) \times (d_2^2-1)$ matrix.

\begin{lemma} Let $U \in SU(d)$, $\lambda_i$ be the generator of $SU(d)$. Then
\begin{equation}
U\lambda_iU^\dag=\sum_{j=1}^{d^2-1}{O_{ij}\lambda_j}
\end{equation}
hold for $i=1,2,\cdots,d^2-1$, where the matrix $(O_{ij})\in SO(d^2-1).$
\end{lemma}

{\bf{Proof:}}  Firstly,
\begin{equation*}
(U\lambda_iU^\dag)^\dag=U\lambda_i^\dag
U^\dag=\sum{\bar{O}_{ij}\lambda_j^\dag}=\sum{\bar{O}_{ij}\lambda_j}.
\end{equation*}
Secondly,
\begin{equation*}
(U\lambda_iU^\dag)^\dag=U\lambda_i^\dag
U^\dag=U\lambda_iU^\dag=\sum{O_{ij}\lambda_i}.
\end{equation*}
These two equations deduce to $O_{ij}=\bar{O}_{ij}$ and $O_{ij}\in R$, where $\bar{O}_{ij}$ means conjugate of $O_{ij}$.

Then due to
\begin{equation*}
tr(\lambda_i\lambda_j)=2\delta_{ij},
\end{equation*}
it has
\begin{equation*}
tr((U\lambda_iU^\dag)(U\lambda_jU^\dag))=tr(U\lambda_i\lambda_j
U^\dag)=tr(\lambda_i\lambda_j)=2\delta_{ij}.
\end{equation*}
Since $\{\lambda_i\}$ is an orthogonal basis of traceless Hermitian matrix of order $d$, so
${U\lambda_i\lambda_j U^\dag}$ is also an orthogonal basis.

Finally, we can get that
\begin{equation}
(U\lambda_1 U^\dag,\cdots,U\lambda_{d^2-1} U^\dag )
=(\lambda_1,\cdots,\lambda_{d^2-1})O^t,
\end{equation}
where $O^t$ is a transitional matrix between the orthogonal bases
${\lambda_i}$ and ${U\lambda_i\lambda_j U^\dag}$, $O^t$ is the transposition of $O$. So $O^t$ and $O$ are real
orthogonal matrices that ends the proving. \qed

\begin{lemma} Two mixed states $\rho$ and $\rho'$ are local unitary
equivalent if and only if there are special orthogonal matrices $O^1\in SO(d_1^2-1)$, $O^2\in SO(d_2^2-1)$ such that
\begin{equation}
S^{\prime}=(O^1)^{t} S,\ T^{\prime}=(O^2)^{t} T,\ R^{\prime}= (O^1)^{t} R O^2.
\end{equation}
\end{lemma}

{\bf{Proof:}}
\begin{eqnarray*}\label{ghh}
\rho^{\prime}&&=(U_1\otimes{U_2})\rho(U_1\otimes{U_2})^\dag\\
&&=\frac{1}{d_1d_2} I\otimes{I}+{\sum_{i=1}}S_{i}(U_1\lambda^1_iU_1^\dagger\otimes
I) +{\sum_{i=1}}T_i (I\otimes U_2\lambda^2_iU_2^\dagger)
\\
&&+{\sum_{i,j=1}}R_{ij}(U_1\lambda^1_iU_1^\dagger)\otimes (U_2\lambda^2_jU_2^\dagger)\\
&&=\frac{1}{d_1d_2} I\otimes{I}+{\sum_{i=1}\sum_{j=1}} S_{i}O^1_{ij} \lambda^1_j \otimes
I +{\sum_{i=1}\sum_{j=1}} T_iO^2_{ij} I\otimes \lambda^2_j
\\
&&+{\sum_{i,j=1}} R_{ij}O^1_{ik}O^2_{jl} \lambda^1_k \otimes \lambda^2_l
\end{eqnarray*}

Comparing the end of above equality and Eq. (\ref{rhoprime}), we can obtain that $S^{\prime}=(O^1)^{t} S$. Similarly, we also arrive
at $T^{\prime}=(O^2)^{t} T$ and
$ R^{\prime}= (O^1)^{t} R O_2$, where $O^1\in SO(d_1^2-1)$, $O^2\in SO(d_2^2-1)$. \qed

Based on the coefficient matrices $S$, $R$, and $T$, we can construct the following invariants for quantum states under the local unitary equivalence.

\begin{theorem}
If two states $\rho$ and $\rho^\prime$ are local unitary equivalent, then they have the same values for the following four sets of invariant:
\begin{eqnarray}\label{jkk}
\begin{array}{rcl}
&&(i)\ S^t(RR^t)^\alpha S, S^t(RR^t)^\alpha RT,\quad \alpha=0,1,\cdots,d_1^2-2;\\
&&(ii)\  T^t(R^tR)^\alpha T,\quad \alpha=0,1,\cdots,d_2^2-2;\\
&&(iii)\  tr(RR^t)^\beta,\quad \beta=1,2,\cdots,d_1^2-1;\\
&&(iv)\  \det R\  \text{when}\  d_1=d_2.
\end{array}
\end{eqnarray}
\end{theorem}

\textbf{Proof:}
First we construct two sets of vectors,
\begin{equation}
\begin{aligned}
&\Omega_1=\left\{S,RT,RR^tS,RR^tRT,(RR^t)^2S, (RR^t)^2RT, \cdots\right\},
 \\
&\Omega_2=\left\{T,R^tS,R^tRT,R^tRR^tS, (R^tR)^2T,(R^tR)^2R^tS, \cdots\right\}.
\end{aligned}
\end{equation}
The first set is consisted by $d_1^2-1$ dimensional vectors and the second set is consisted by $d_2^2-1$ dimensional vectors. Although there are infinite elements in sets $\Omega_i$, there are at most $d_i^2-1$ linear independent vectors, $i=1,2$.
Under local unitary transformations, one can verify the inner product of two elements in $\Omega_i$ is invariant, that is, $\left\langle\mu_k,\mu_l\right\rangle=tr \mu_k^\dagger \mu_l$ is invariant for $\mu_k \in \Omega_i$, $i=1,2$. Employing these two sets $\Omega_1$ and $\Omega_2$, we get $S^t(RR^t)^\alpha S$, $S^t(RR^t)^\alpha RT$, $T(R^tR)^\alpha T$ are invariants.

Next, we only need to determine the scope of $\alpha$. Because $RR^t$ is a $(d_1^2-1)\times (d_1^2-1)$ matrix, by Cayley-Hamilton theorem we can arrive at that
\begin{equation}\nonumber
(RR^t)^{d_1^2-1}=e_1(RR^t)^{d_1^2-2}-e_2(RR^t)^{d_1^2-3}+\cdots+e_{d_1^2-2}(RR^t)-e_{d_1^2-1}I,
\end{equation}
where $e_i$ is a polynomial about the eigenvalues of $RR^t$. Hence $(RR^t)^{d_1^2-1}$ is a
linear combination of $(RR^t)^{d_1^2-2},(RR^t)^{d_1^2-3},\cdots,(RR^t),I$. So there
are at most ${d_1^2-1}$ linear independent vectors in the sets of $\{(RR^t)^{\alpha}\}$. Then we have
$\alpha=0,1,2,\cdots,d_1^2-2$ for the first set of invariants. Similarly, it is easy to obtain the scope of powers of the second set of invariants.

Now we check the third set of invariants. Let $\mu_i$ be the
eigenvalue of $RR^t$ and $\mu_i^{\prime}$ be the eigenvalue of
$R^{\prime}R^{\prime t}$, then
$tr(RR^t)^\beta=\sum_{i=1}^{d_1^2-1}{\mu_i^\beta}$. We all know that the
characteristic polynomial of $RR^t$ is $\mid{\mu_i I-RR^t}\mid$,
\begin{gather}\nonumber
\begin{aligned}
& \quad\mid{\mu_i^{\prime} I-R^{\prime}R^{\prime t}}\mid\\
&=\mid{\mu_i^{\prime} I-(O^1)^tRO^2(O^2)^tR^tO^1}\mid\\
&=\mid{\mu_i^{\prime} I-(O^1)^t(RR^t)O^1}\mid\\
&=\mid{(O^1)^t(\mu_i^{\prime} I-RR^t)O^1}\mid\\
&=\mid (O^1)^t\mid\mid{\mu_i^{\prime} I-RR^t}\mid\mid O^1\mid\\
&=\mid{(O^1)^tO^1}\mid\mid{\mu_i^{\prime}
I-RR^t}\mid\\
&=\mid{\mu_i^{\prime} I-RR^t}\mid.
\end{aligned}
\end{gather}
so $\mu_i=\mu_i^{\prime}$, which is the singular value of $R$. The invariance of singular values of $R$ is equivalent to the invariance of $tr(RR^t)^\beta$.
Next we need to determine the scope of $\beta$. Because $RR^t$ is a
$(d_1^2-1)\times (d_1^2-1)$ matrix, there are at most $d_1^2-1$ nonzero eigenvalues. Therefore we
need at most $d_1^2-1$ equations about $\mu_i$, then $\beta$ is at most
equal to $d_1^2-1$.

At last, when $d_1=d_2$, $R$ is a square matrix. Then $\det R^{\prime}=\det((O^1)^t R O^2)=\det (O^1)^t \det R \det O^2$ with
$O^1$, $O^2$ $\in SO(d_1^2-1)$, and $\det (O^1)^t=\det O^2=1$. Therefore
$\det R^{\prime}=\det R$.\qed

\section{LU Invariants of Three Partite Quantum States}

Now, we consider three partite systems in a $d_1 \times d_2 \times
d_3$ dimensional Hilbert space $H_A \otimes H_B \otimes H_C$. Two
three partite states $\rho$ and $\rho'$ are said to be local unitary
equivalent if there exist unitary operators $U_1\in{SU(d_1)}$,
$U_2\in{SU(d_2)}$ and $U_3\in{SU(d_3)}$ such that
\begin{equation}\label{ass}
\rho^\prime=(U_1\otimes{U_2} \otimes{U_3})\rho(U_1\otimes{U_2} \otimes{U_3})^\dag.
\end{equation}

In general, mixed states $\rho$ and $\rho^\prime$ acting on $H_A \otimes H_B \otimes H_C$ can be decomposed in the following way:
\begin{equation}
\rho=\frac{1}{d_1d_2d_3}I \otimes I \otimes I + \sum_i \sum_m
S^{m}_i \lambda^m_i\otimes I \otimes I + \sum_{i,j} \sum_{m,n}
T^{mn}_{ij} \lambda^m_i \otimes \lambda^n_j \otimes I+ \sum_{ijk}
R_{ijk}\lambda^1_i \otimes \lambda^2_j \otimes \lambda^3_k,
\end{equation}
\begin{equation}
\rho^\prime=\frac{1}{d_1d_2d_3}I \otimes I \otimes I + \sum_i \sum_m
S^{^\prime m}_i \lambda^m_i\otimes I \otimes I + \sum_{i,j}
\sum_{m,n} T^{^\prime mn}_{ij} \lambda^m_i \otimes \lambda^n_j
\otimes I+ \sum_{ijk} R^\prime_{ijk} \lambda^1_i \otimes \lambda^2_j
\otimes \lambda^3_k,
\end{equation}
where $S^{(\prime) m}_i=Tr(\rho^{(\prime)}\lambda^m_i\otimes I\otimes I)$,
$T^{(\prime) mn}_{ij}=Tr(\rho^{(\prime)} \lambda^m_j \otimes
\lambda^n_j \otimes I)$, $R^{(\prime)}_{ijk}=Tr(\rho^{(\prime)}
\lambda^1_i \otimes \lambda^2_j \otimes \lambda^3_k)$,
$\lambda^m_i$ the generator of $SU(d_m)$, $i=1,2,\cdots,d_m^2-1$, $m=1,2,3$.
Here $S^{(\prime)m}$ is a $d_m^2-1$ dimensional vector, $T^{(\prime)mn}$ is a
$(d_m^2-1)\times (d_n^2-1)$ dimensional vector, $m,n=1,2,3$. $R$ is a
hypermatrix. If we regard its first subscript as the row index and
the other two as column index, then $R$ can be written as
$R_{1|23}$. If we regard the second subscript as the row index and
the other two as column index, then $R$ can be written as
$R_{2|13}$. At last, if we regard the third subscript as the row
index and the other two as column index, then $R$ can be written as
$R_{3|12}$.

By the method which is similar to the proof of Lemma 2, we can obtain
\begin{eqnarray*}
S_1^{\prime}&=&(O^1)^t S_1,\\
S_2^{\prime}&=&(O^2)^t S_2,\\
S_3^{\prime}&=&(O^3)^t S_3,\\
T_{12}^{\prime}&=&(O^1\otimes O^2)^t T_{12} ,\\
T_{13}^{\prime}&=&(O^1\otimes O^3)^t T_{13} ,\\
T_{23}^{\prime}&=&(O ^2\otimes O^3)^t T_{23} ,\\
R_{1|23}^{\prime}&=&(O^1)^tR_{1|23}(O^2\otimes O^3),\\
R_{2|13}^{\prime}&=&(O^2)^tR_{2|13}(O^1\otimes O^3),\\
R_{3|12}^{\prime}&=&(O^3)^tR_{3|12}(O^1\otimes O^2).
\end{eqnarray*}
Based on these relations of coefficient matrices of $\rho$ and $\rho^\prime$, we first define six sets of matrices as follows.
\begin{gather}
\begin{aligned}\nonumber
&\left\langle\Omega_1\right\rangle=\left\{S_1,R_{1|23}T_{23},R_{1|23}R_{1|23}^tS_1,R_{1|23}R_{1|23}^tR_{1|23}T_{23},
\cdots\right\},\\
&\left\langle\Omega_2\right\rangle=\left\{S_2,R_{2|31}T_{31},R_{2|31}R_{2|31}^tS_2,R_{2|31}R_{2|31}^tR_{2|31}T_{31},
\cdots\right\},\\
&\left\langle\Omega_3\right\rangle=\left\{S_3,R_{3|12}T_{12},R_{3|12}R_{3|12}^tS_3,R_{3|12}R_{3|12}^tR_{3|12}T_{12},
\cdots\right\},\\
&\left\langle\Omega_{2}\otimes\Omega_{3}\right\rangle_{1|23}=\left\{T_{23},R_{1|23}^tS_1,R_{1|23}^tR_{1|23}T_{23},R_{1|23}^tR_{1|23}R_{1|23}^tS_1,\cdots\right\},\\
&\left\langle\Omega_{1}\otimes\Omega_3\right\rangle_{2|13}=\left\{T_{13},R_{2|13}^tS_2,R_{2|13}^tR_{2|13}T_{13},R_{2|13}^tR_{2|13}R_{2|13}^tS_2,\cdots\right\},\\
&\left\langle\Omega_1\otimes \Omega_2\right\rangle_{3|12}=\left\{T_{12},R_{3|12}^tS_3,R_{3|12}^tR_{3|12}T_{12},R_{3|12}^tR_{3|12}R_{3|12}^tS_3,\cdots\right\}.
\end{aligned}
\end{gather}
Here the set $\Omega_m$ is consisted by $d_m^2-1$ dimensional vectors and $\Omega_m\otimes \Omega_n$ is consisted by $(d_m^2-1)\times (d_n^2-1)$ dimensional vectors, $m,n=1,2,3$.

\begin{theorem}
If two three partite quantum states are local unitary equivalent,
then they have the same values of the following invariant:
\begin{equation}
\begin{array}{rcl}
(i)&&S_1^t(R_{1|23}R_{1|23}^t)^{\alpha_1}S_1,\ S_1^t(R_{1|23}R_{1|23}^t)^{\alpha_1} R_{1|23}T_{23},\ \alpha_1=0,1,\cdots,d_1^2-2,\\
&&S_2^t(R_{2|13}R_{2|13}^t)^{\alpha_2}S_2,\ S_2^t(R_{2|13}R_{2|13}^t)^{\alpha_2}R_{2|13}T_{13},\ \alpha_2=0,1,\cdots,d_2^2-2,\\
&&S_3^t(R_{3|12}R_{3|12}^t)^{\alpha_3}S_3,\ S_3^t(R_{3|12}R_{3|12}^t)^{\alpha_3}R_{3|12}T_{12},\ \alpha_3=0,1,\cdots,d_3^2-2,\\
&&T_{12}^t(R_{3|12}^tR_{3|12})^{\alpha_{12}}T_{12},\ \alpha_{12}=0,1,\cdots,(d_1^2-1)\times (d_2^2-1)-2,\\
&&T_{23}^t(R_{1|23}^tR_{1|23})^{\alpha_{23}}T_{23}, \ \alpha_{23}=0,1,\cdots,(d_2^2-1)\times (d_3^2-1)-2,\\
&&T_{13}^t(R_{2|13}^tR_{2|13})^{\alpha_{13}}T_{13},\ \alpha_{13}=0,1,\cdots,(d_1^2-1)\times (d_3^2-1)-2;\\
(ii)&&tr(R_{1|23} R_{1|23}^t )^{\beta_1},\ \beta_1=1,2,\cdots,d_1^2-1,\\
&&tr(R_{2|13} R_{2|13}^t )^{\beta_2},\ \beta_2=1,2,\cdots,d_2^2-1,\\
&&tr(R_{3|12} R_{3|12}^t)^{\beta_3},\ \beta_2=1,2,\cdots,d_2^2-1,\\
&&tr(T_{12}T_{12}^t)^{\beta_{12}},\ \beta_{12}=1,2,\cdots,(d_1^2-1)\times(d_2^2-1)-1,\\
&&tr(T_{23}T_{23}^t)^{\beta_{23}},\ \beta_{23}=1,2,\cdots,(d_2^2-1)\times(d_3^2-1)-1,\\
&&tr(T_{13}T_{13}^t)^{\beta_{13}},\ \beta_{13}=1,2,\cdots,(d_1^2-1)\times(d_3^2-1)-1.
\end{array}
\end{equation}
\end{theorem}

{\bf{Proof:}}
For the first set of quantities, one can verify they are invariant under local unitary transformations considering the invariance of inner product of two elements in $\Omega_m$ and $\Omega_m\otimes \Omega_n$, $m,n=1,2,3$.
For example,
\begin{gather}\nonumber
\begin{aligned}
&{S_1^t}^{\prime}({R_{1|23}}^{\prime}({R_{1|23}^{\prime t}}))^{\alpha_1}{S_1}^{\prime}\\
=&S_1^tO^1((O^1)^tR_{1|23}(O^2\otimes O^3)(O^2\otimes O^3)^tR_{1|23}^tO^1)^{\alpha_1}(O^1)^tS_1;\\
=&{S_1^t}({R_{1|23}}({R_{1|23}^{t}}))^{\alpha_1}{S_1}
\\
&{S_1^t}^{\prime}{(R_{1|23}}^{\prime}{R_{1|23}^{\prime t}})^{\alpha_1}{R_{1|23}}^{\prime}T_{23}^{\prime}\\
=&S_1^tO^1((O^1)^tR_{1|23}(O^2\otimes O^3)(O^2\otimes O^3)^tR_{1|23}^tO^1)^{\alpha_1}(O^1)^tR_{1|23}(O^2\otimes O^3)(O^2\otimes O^3)^tT_{23}\\
=&S_1^t(R_{1|23}R_{1|23}^t)^{\alpha_1} R_{1|23}T_{23};\\
\end{aligned}
\end{gather}
The scope of powers $\alpha_m$ and $\alpha_{mn}$ can be easily derived by the orders of the coefficient matrices and Cayley-Hamilton theorem.  \qed

In three partite systems, the coefficient matrices $T$ and $R$ can
be represented in more than one way. For example, we can regard
$T_{12}$ as a vector or a matrix with the first subscript as the row
index and the second as the column index. More than that, we can
also regard $R_{123}$ as a matrix or a large dimensional vector. The
transformations of these matrices under local unitary
transformations are related for different representations. For
coefficient matrix $T_{12}$, if we regard it as a vector, then under
local unitary transformations it is changed to $(O^1\otimes O^2)^t
T_{12}$. If we regard it as a matrix with the first subscript as the
row index and the second as the column index, then it is changed to
$(O^1)^t T_{12} O^2$. Now we regard $T_{mn}$ as matrix instead of
vector and get the following invariants.

\begin{theorem}
If two three partite quantum states are local unitary equivalent,
then they have the same values of the following invariant:
\begin{equation}
\begin{array}{rcl}
(i)&&S_1^t(T_{12}T_{12}^t)^{\alpha_1}S_1,\ \alpha_1=0,1,\cdots,(d_1^2-1)\times(d_1^2-1)-2,\\
&&S_2^t(T_{23}T_{23}^t)^{\alpha_2}S_2,\ \alpha_2=0,1,\cdots,(d_2^2-1)\times(d_2^2-1)-2,\\
&&S_3^t(T_{31}T_{31}^t)^{\alpha_3}S_3,\ \alpha_3=0,1,\cdots,(d_3^2-1)\times(d_m^2-1)-2;\\
(ii)&&\det T_{ij} \ \text{when $d_i$=$d_j$}, i,j=1,2,3.
\end{array}
\end{equation}
\end{theorem}

{\bf{Proof:}} When we regard $T_{mn}$ as matrix instead of vector, we know that
$T_{12}^{\prime}=O_1^tT_{12}O_2$, $T_{31}^{\prime}=O_3^tT_{31}O_1$, $T_{23}^{\prime}=O_2^tT_{23}O_3$.
So it is easy to verify the above two conclusions.

\section{Conclusion and discussions}
We studied the local unitary equivalence of quantum states in terms of invariants. In bipartite system, we expand quantum states in Bloch representation first. Then some invariants under local unitary transformation are constructed by the products, the singular values and the determinant of coefficient matrix. Similarly, we get the invariants for three partite system. This method and result can be generalized to high dimensional multipartite system.

\bigskip
\noindent{\bf Acknowledgments}\ The work is partly supported by the
NSF of China under Grant No. 11501153 and No. 11401032; the NSF of Hainan
Province under Grant No. 20161006.

\end{document}